**Tuning the intermediate reaction barriers by CuPd catalyst to improve the selectivity of electroreduction CO$_2$ to C2 products**

Li Zhu [a], Yiyang Lin [a], Kang Liu [a], Emiliano Cortés [b], Hongmei Li [a], Junhua Hu [c], Akira Yamaguchi [d], Ying-Rui Lu [e], Ting-Shan Chan [e], Xiaoliang Liu [a,*], Masahiro Miyauchi [d,*], Junwei Fu [a,*], and Min Liu [a,*]

[a] *Shenzhen Research Institute, School of Physics and Electronics, Central South University, P. R. China.*

*E-mail: minliu@csu.edu.cn, xl_liu@csu.edu.cn, fujunwei@csu.edu.cn*

[b] *Chair in Hybrid Nanosystems, Nanoinstitute Munich, Faculty of Physics, Ludwig-Maximilians-Universität München, 80539 München, Germany.*

[c] *School of Materials Science and Engineering, Zhengzhou University, China.*

[d] *Department of Materials Science and Engineering, School of Materials and Chemical Technology, Tokyo Institute of Technology, Tokyo 152-8552, Japan. E-mail: mmiyauchi@ceram.titech.ac.jp*

[e] *National Synchrotron Radiation Research Center, Hsinchu 300, Taiwan.*

**Abstract**

Electrochemical CO$_2$ reduction is a promising strategy for utilization of CO$_2$ and intermittent excess electricity. Cu is the only single-metal catalyst that can electrochemically convert $CO_2$ to multi-carbon products. However, Cu has an undesirable selectivity and activity for C2 products, due to its insufficient amount of CO* for C-C coupling. Considering the strong $CO_2$ adsorption and ultra-fast reaction kinetics of CO* formation on Pd, an intimate CuPd(100) interface was designed to lower the intermediate reaction barriers and then improve the efficiency of C2 products. Density functional theory (DFT) calculations showed that the CuPd(100) interface has enhanced $CO_2$ adsorption and decreased $CO_2$* hydrogenation energy barrier, which are beneficial for C-C coupling. The potential-determining step (PDS) barrier of $CO_2$ to C2 products on CuPd(100) interface is 0.61 eV, which is lower than that on Cu(100) (0.72 eV). Motivated by the DFT calculation, the CuPd(100) interface catalyst was prepared by a facile chemical solution method and demonstrated by transmission electron microscope (TEM). The $CO_2$ temperature programmed desorption ($CO_2$-TPD) and gas sensor experiments proved the enhancements of $CO_2$ adsorption and $CO_2$* hydrogenation abilities on CuPd(100) interface catalyst. As a result, the obtained CuPd(100) interface catalyst exhibits a C2 Faradaic efficiency of 50.3±1.2% at −1.4 $V_{RHE}$ in 0.1 M $KHCO_3$, which is 2.1 times higher than 23.6±1.5% of Cu catalyst. This work provides a rational design of Cu-based electrocatalyst for multi-carbon products by fine-tuning the intermediate reaction barriers.

*Keywords:* carbon dioxide reduction; C2 chemicals; Electrocatalyst; Copper-palladium interface; Intermediate reaction barriers.

## 1. Introduction

Excessive carbon emissions have caused serious global environmental issues.[1-3] Using intermittent excess electricity to electrochemically convert $CO_2$ to valuable chemicals is a potential strategy to simultaneously solve the Earth's carbon recycle and energy crisis.[4-8] Among various $CO_2$ reduction products, C2 products (e.g., $C_2H_4$, $C_2H_5OH$) have attracted much attention due to their higher energy density compared with that of C1 products (e.g., $HCOOH$, $CH_4$, $CH_3OH$).[9-11] Cu is the unique single-metal catalyst to electrochemically reduce $CO_2$ into multi-carbon ($C_{2+}$) products.[12-14] However, pure Cu catalysts do not exhibit desirable selectivity and activity of C2 products for practical application.[15,16] How to improve the efficiency of C2 products on Cu and Cu-based catalysts has aroused great interest.[17-20]

There are two limiting parts for electroreduction $CO_2$ to C2 products: the amount of CO* as the carbon source (* represents the adsorbates on the surface of substrate),[21] and the C-C coupling step (two adjacent CO* coupling).[22,23] For Cu catalysts, the energy barrier of C-C coupling step is relatively low.[24,25] However, the abilities of $CO_2$ adsorption and $CO_2$* hydrogenation on Cu are undesirable,[26,27] resulting in insufficient amount of adsorbed CO*. As such, different approaches were explored to improve the catalytic activity of Cu for C2 products.[28-30] Among them, the design of Cu-based bimetallic catalyst is one of the most promising strategies.[31-33] In principle, the second metal component can effectively adjust the binding energy between catalyst and intermediates,[34-36] lower the energy barriers of intermediate reaction, and then increase C2 products efficiency.[37,38] Palladium (Pd), as an efficient catalyst, exhibits strong $CO_2$ adsorption and ultra-fast reaction kinetics for CO* formation. However, CO* poisoning on Pd surface makes it not suitable for generating C2 products.[39] In order to take full advantages of both Cu (C-C coupling) and Pd (CO* formation), constructing a Cu-Pd bimetallic catalyst is a potential method to optimize the efficiency of C2 products formation.

In this work, we constructed a CuPd (100) interface catalyst to tune the intermediate reaction barriers and improve the C2 products selectivity. Density functional theory (DFT) calculations predicted that the CuPd(100) interface is able to strongly adsorb $CO_2$ and dramatically decrease the energy barrier of $CO_2^*$ hydrogenation compared with those of Cu(100) facet, leading to sufficient CO* for the later C-C coupling step. The calculated potential-determining step (PDS) of $CO_2$ to C2 products on CuPd(100) interface is C-C coupling and the corresponding energy barrier is 0.61 eV, which is much lower than that of 0.72 eV on Cu(100) for the PDS ($CO_2^*$ hydrogenation), indicating a potential higher efficiency of C2 products on the CuPd(100) interface catalyst. Experimentally, the CuPd (100) interface catalyst was prepared by an in-situ growth method using thermal-reduction to obtain Pd nanoparticles (NPs) as nucleation seeds. The obtained CuPd(100) interface catalyst was assessed by XRD, TEM and XPS. The enhancements of $CO_2$ adsorption and $CO_2^*$ hydrogenation abilities on CuPd(100) interface were proved by $CO_2$-TPD and gas sensor experiments, respectively. As a result, the CuPd(100) interface catalyst exhibits a C2 Faradaic efficiency (FE) of 50.3±1.2% at −1.4 $V_{RHE}$ in 0.1 M $KHCO_3$, which is 2.1 times higher than Cu catalyst (23.6±1.5%). This work provides a strategy to improve the yield of target C2 products by regulating the energy barrier of intermediate reaction, and a reference for the development of Cu-based catalysts with higher efficiency of multi-carbon products.

## 2. Experimental sections

### 2.1 DFT calculations

To explore the mechanisms of $CO_2$ to C2 products, $4 \times 2$ Cu(100), Pd(100) and CuPd(100) periodic surface slab including four atomic layers were built as shown in Fig. S1. The main consideration is that the Cu(100) facet favours the formation of C2 products.[40,41] A vacuum slab with 30 Å was added to avoid the interaction influence of the periodic boundary conditions. Each model contains 128 atoms. The potassium (K) ions not only promote the activation of $CO_2$,[42] but also lowers the energy barrier of C-C coupling.[24] Six K ions were added on the model to simulate the actual $CO_2$ reduction process.

The DFT calculations were performed by VASP with the projector augment wave (PAW) method.[43,44] The exchange and correlation potentials are present in the generalized gradient approximation with the Perdewe-Burkee-Ernzerhof (GGA-PBE).[45,46] A $2 \times 2 \times 1$ gamma grid of k-points was used for the Brillouin zone integration. The cutoff energy, the convergence criteria for energy and force were set as 450 eV, $10^{-5}$ eV/atom and 0.02 eV/Å, respectively.

The adsorption energy was calculated by the following formula:[47,48]

$$E_{ads} = E_{substrate+gas} - (E_{substrate} + E_{gas})$$

Where $E_{substrate}$ and $E_{gas}$ represent the energy of the isolated substrate and gas molecule, respectively. The $E_{substrate+gas}$ represents the total energy of gas molecule adsorbed on the substrate. Herein, the substrates refer to Cu(100), Pd(100) and CuPd(100) interface.

The change of Gibbs free energy ($\Delta G$) for each reaction step is given as follow:[49,50]

$$\Delta G = \Delta E + \Delta ZPE - T\Delta S$$

Where $\Delta E$ represents the total energy difference between the product and the reactant. $\Delta ZPE$ and $T\Delta S$ are the zero-point energy correction and the entropy change at 298.15 K, respectively.

2.2 Catalyst synthesis

Preparation of Cu sample: 3 mmol of copper acetate was thoroughly dissolved into 250 mL of 2-ethoxyethanol with vigorous stirring and Ar bubbling. After 30 min, 20 mL of $NaBH_4$ aqueous solution (1.5 M) was dropwise added into the above solution. The obtained black precipitate was washed with water and ethanol for several times. The collected Cu sample was dried at 60 $^{o}$C for 6 h in vacuum, and then dispersed in isopropyl alcohol.[51]

Preparation of Pd sample: 3 mmol of palladium acetate was firstly dissolved into 30 mL of acetone, then 250 mL of 2-ethoxyethanol was added. Stir evenly, 20 mL of $NaBH_4$ aqueous solution

(1.5 M) was dropwise added into the mixture solution. The obtained black precipitate was washed with water and ethanol for several times, and dried at 60 °C for 6 h in vacuum, then dispersed in isopropyl alcohol.

Preparation of CuPd sample: 1.5 mmol of palladium acetate was dissolved into 10 mL of acetone. 250 mL of 2-ethoxyethanol was added and heated to 393 K for 30 min with vigorous stirring and Ar bubbling. After cooling to room temperature, 20 mL of copper acetate aqueous solution (75 mmol/L) was dropwise added with stirring, and followed by 20 mL of $NaBH_4$ aqueous solution (1.5 M). The obtained black precipitate was thoroughly washed with water and ethanol, and dried at 60 °C for 6 h in vacuum, then dispersed in isopropyl alcohol.

## 3. Results and discussion

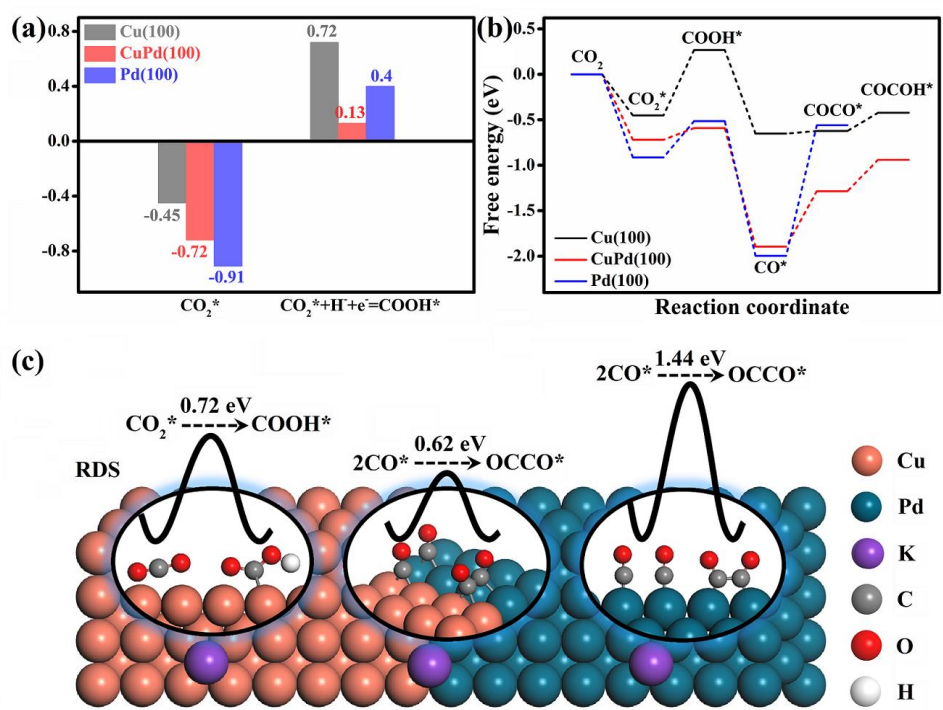

**Fig. 1** (a) The adsorption energy of $CO_2$ and the $\Delta G$ of $CO_2$* hydrogenation, (b) calculated free energy diagrams of $CO_2$ reduction process, (c) schematic diagram of the PDS of $CO_2$ reduction process on Cu(100), CuPd(100) interface and Pd(100) facets.

The adsorption energy of $CO_2$ and the $\Delta G$ of $CO_2^*$ hydrogenation are shown in Fig. 1a and Fig. 1b.[19,36,52-54] The adsorption energy of $CO_2$ on CuPd(100) interface (−0.72 eV) is stronger than that on Cu(100) (−0.45 eV) and weaker than that on Pd(100) (−0.91 eV). Fig. S3 shows the adsorption of $CO_2$ on these models without $K^+$ are much weaker than that with $K^+$. We can infer that the $CO_2$ adsorption is strongly increased by the $K^+$.[42] The $\Delta G$ of $CO_2^*$ hydrogenation on CuPd(100) interface is greatly decreased to 0.13 eV from 0.72 eV on Cu(100), and even lower than that on Pd(100) (0.40 eV). Considering the fact that the zero-point energy and the entropy energy correction are extremely small, the $\Delta G$ of $CO_2^*$ hydrogenation depends mainly on the difference between $E_{\text{substrate+COOH*}}$ and $E_{\text{substrate+CO2*}}$. The more negative the $E_{\text{substrate+COOH*}}$ (the stronger COOH* adsorption on substrate) and the more positive the $E_{\text{substrate+CO2*}}$ (the weaker $CO_2^*$ adsorption on substrate), the smaller the $\Delta G$ of $CO_2^*$ hydrogenation. Therefore, the dramatic decrease in the $\Delta G$ of $CO_2^*$ hydrogenation on CuPd(100) interface can be attributed to the strong COOH* adsorption and appropriate $CO_2$ adsorption of CuPd(100) interface (Fig. S2 and S3).

Fig. S5a shows the adsorption energy of CO* on Cu(100), CuPd(100) interface and Pd(100) facet with K ions are −1.2, −2 and −2.4 eV, respectively. Thus, more sufficient CO* will be formed on CuPd(100) interface compared with the Cu (100) facet, leading to higher chance for the C-C coupling step. In addition, the adsorption energies of CO* with different CO* coverage on Pd(100) facet in Fig. S5b. The results show that the adsorption of CO* on Pd(100) become weaker as the CO* coverage increasing, which indicate that the CO* desorption from Pd will be possible with high CO* coverage.[55]

Fig. S6 shows the energy barriers of two CO* coupling on Cu(100) and CuPd(100) are less than that of CO* hydrogenation. Therefore, it is considered that C-C coupling step on Cu(100) and CuPd(100) involve two CO* coupling, not COH*/CHO* coupling. The obtained free energy of two CO* coupling on Cu(100), CuPd(100) and Pd(100) facet are 0.09, 0.61 and 1.44 eV, respectively

(Fig. 1b). The large $\Delta G$ on Pd(100) facet is consistent with the previous reported results, proving that Pd has no catalytic activity for C2 products generation.

As previously reported, the PDS step of $CO_2$ electroreduction to C2 products is $CO_2$ activation or C-C coupling,[40] therefore, it is considered that the PDS of $CO_2$ reduction to C2 products on CuPd(100) interface is C-C coupling with an energy barrier of 0.61 eV, which is lower than 0.72 eV of $CO_2*$ hydrogenation on Cu(100). Therefore, the CuPd(100) interface facilitates the conversion of $CO_2$ to C2 products compared with that of Cu(100) facet.

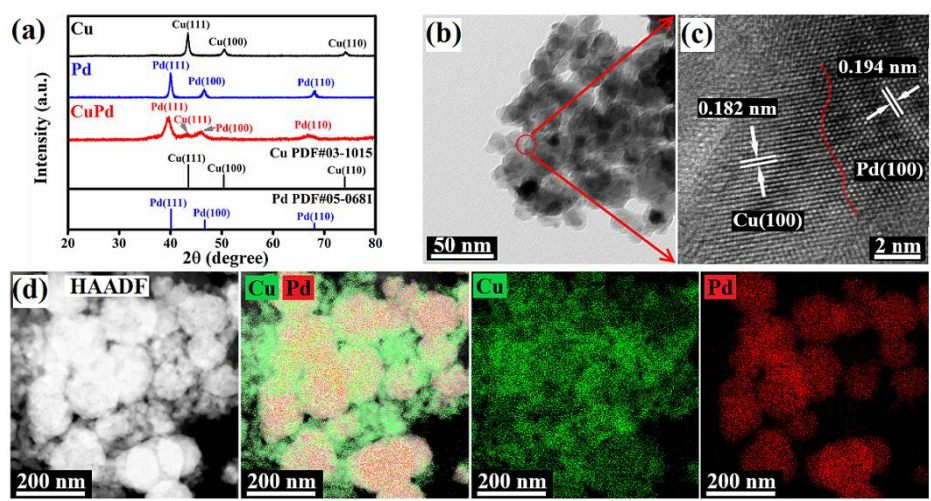

**Fig. 2** (a) XRD pattern of the prepared Cu, CuPd and Pd samples after electrochemical reduction for 30 min. (b, c) Low-resolution and high-resolution TEM images of CuPd sample. (d) HAADF-STEM image, combined with EDS mapping of CuPd sample.

Motivated by these predictions, we prepared Cu, CuPd and Pd samples by thermal-reduction treatment followed with an in-situ growth process. As shown in Fig. S4, the XRD patterns show that the CuPd sample before electrochemical reduction contain the characteristic peaks of Cu (PDF No.03-1015), Pd (PDF No.05-0681) and $Cu_2O$ (PDF No.78-0428). After electrochemical reduction for 30 min, the characteristic peak of $Cu_2O$ disappeared (Fig. 2a). These results reveal that we do not need to consider the effect of oxidation state of Cu on the catalytic activity of these catalysts, and the CuPd sample consists of separate Cu phase and Pd phase rather than CuPd alloy.[51,56] Fig. 2b shows the TEM image of CuPd sample, which has a typical nanoparticle morphology with size of around 20 nm. The high-resolution TEM (HRTEM) image (Fig. 2c) shows lattice distances of 0.182 nm and 0.194 nm, which are corresponding to Cu(100) and Pd(100) facets, respectively.[29] The red line shows the clear CuPd(100) interface. Fig. 2d shows the HAADF-STEM and EDS mapping images, proving the separate distribution of Cu (green) and Pd (red) phases. In addition, the TEM images and HRTEM images of Cu and Pd sample are shown in Fig. S5, the lattice distances of 0.181 nm and 0.195 nm are corresponding to the Cu(100) facets of Cu sample and Pd(100) facets of Pd sample, respectively.

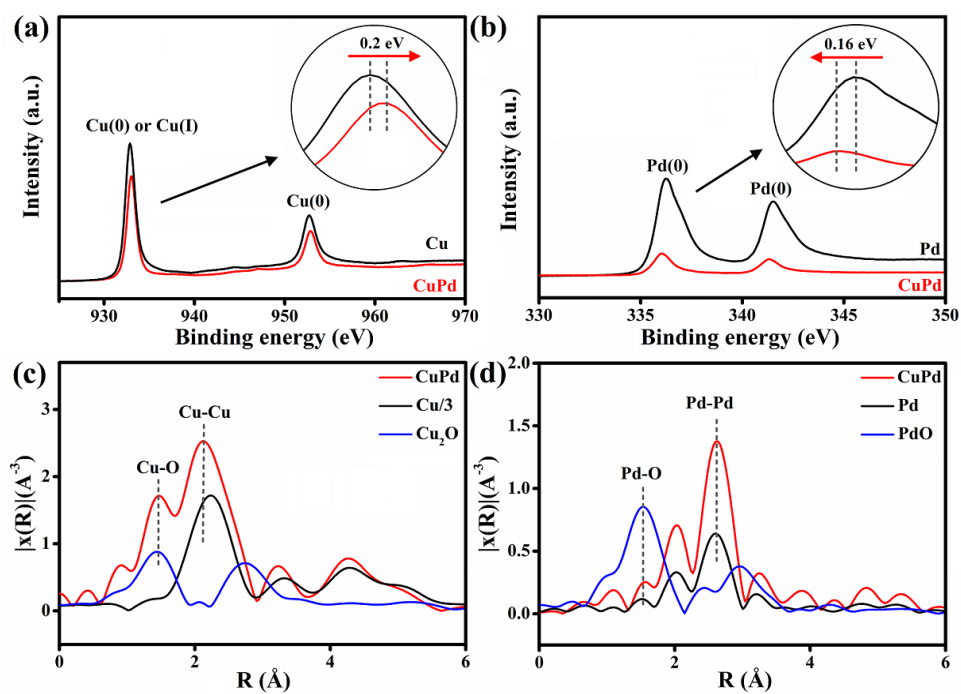

**Fig. 3** XPS spectra of (a) Cu 2p and (b) Pd 3d, EXAFS spectra of (c) Cu K-edge and (d) Pd K-edge of CuPd sample.

XPS was further used to study the composition and elemental chemical state of the samples.[57-59] As shown in Fig. 3a and 3b, the Cu 2p binding energy of CuPd sample exhibits 0.2 eV positive shift compared with Cu sample, and the Pd 3d binding energy of CuPd sample moves 0.16 eV towards the low energy region compared with the Pd sample. The slight shifts of the binding energies of Cu 2p and Pd 3d indicate the electron transfer from Cu to Pd, revealing an intimate interaction between Cu and Pd in CuPd sample.[30] Fig. 3c and 3d show Cu K-edge and Pd K-edge extended X-ray absorption fine structure (EXAFS) spectra of CuPd samples. Only the Cu-Cu bonds and the Pd-Pd bonds can be found in the spectra. XPS and EXAFS results together with XRD and TEM analyses clearly demonstrate that the obtained CuPd catalyst is a phase-separated sample with CuPd(100) interfaces.

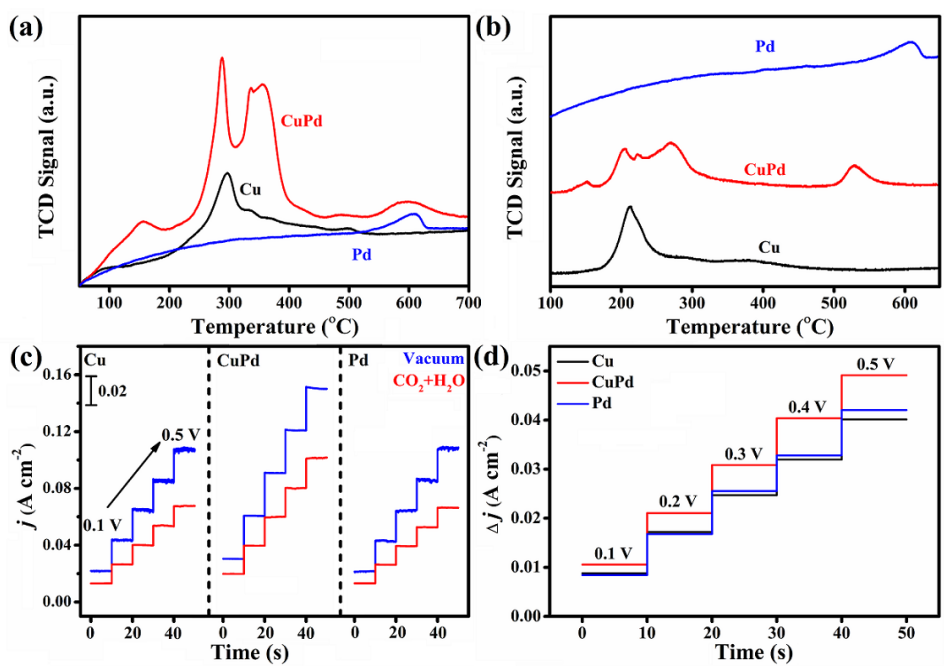

**Fig. 4** (a) The $CO_2$-TPD and (b) CO-TPD curves of Cu, CuPd(100) interface and Pd catalysts. (c) Gas sensor experiments for Cu, CuPd(100) interface and Pd catalysts at 0.1, 0.2, 0.3, 0.4 and 0.5 V. (d) The calculated current density differences ($\Delta j$) between vacuum and $CO_2$+$H_2O$ atmosphere.

To verify the $CO_2$ adsorption ability of the three catalysts, the $CO_2$-TPD measurements and thermo-gravimetric experiments were carried out (Fig. 4a and Fig. S6). The main $CO_2$-desorption peak for the Cu catalyst is located at 296 °C,[60] and the $CO_2$-desorption peak for Pd catalyst is located at 608 °C.[61] For the CuPd(100) interface catalyst, there are three main $CO_2$-desorption peaks located at 288, 355 and 598 °C. Compared with the Cu and Pd catalysts, the peak located at 355 °C can be assigned to the CuPd(100) interface, suggesting the stronger $CO_2$ adsorption than that on Cu and weaker $CO_2$ adsorption than that on Pd.

To prove the strong ability of COOH* adsorption of CuPd(100) interface catalyst, we designed the gas sensor experiment (Fig. S7).[62,63] Fig. 4c shows the current density curves of Cu, CuPd(100) interface and Pd catalysts at different applied potentials in vacuum and saturated $CO_2$+$H_2O$ atmosphere. Fig. 4d shows the calculated current density differences ($\Delta j$) between vacuum and $CO_2$+$H_2O$ atmosphere. The higher the $\Delta j$, the stronger adsorption of $CO_2$ and $H_2O$, which is an important indicator of COOH* adsorption. From these results, it can be observed that the CuPd(100) interface catalyst has the strongest COOH*adsorption ability.

Further, the CO-TPD was used to investigate the CO adsorption ability of these three catalysts.[64,65] As shown in Fig. 4b, CuPd(100) interface catalyst exhibits three obvious desorption peaks located at 204, 271 and 529 $^o$C. As the reference catalysts, Cu and Pd show desorption peaks at 211 and 609 $^o$C, respectively.[66,67] Combined with the three CO desorption curves, the peak at 271 $^o$C can be ascribed to the contribution of CuPd(100) interface. The moderate desorption temperature (271 $^o$C) of CuPd(100) interface compared with Cu (204 $^o$C) and Pd (609 $^o$C) suggests the moderate CO adsorption ability on CuPd(100) interface.

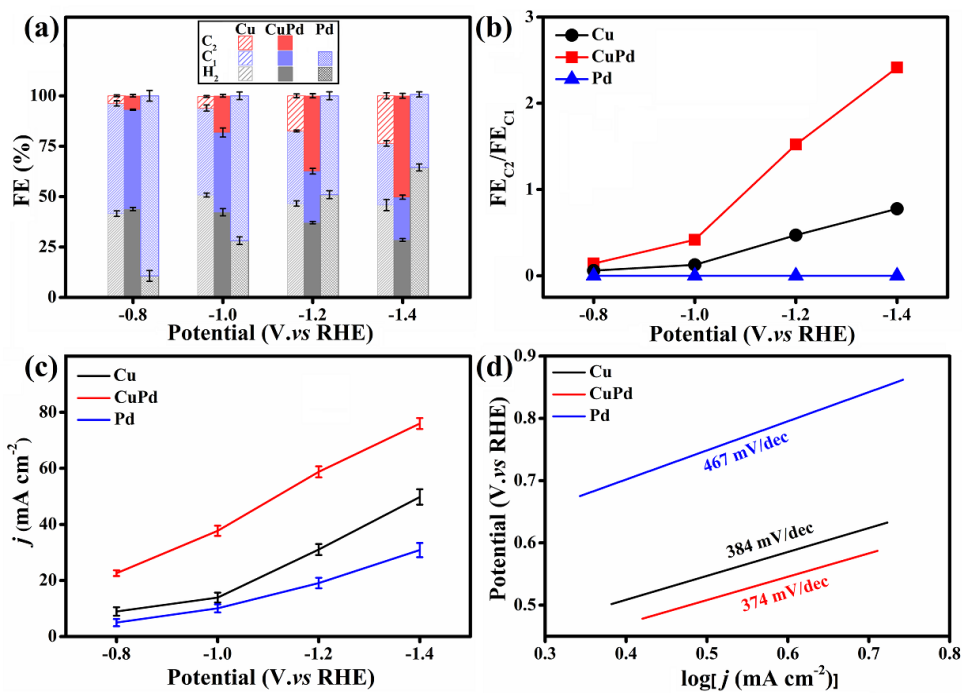

**Fig. 5** (a) The FE of different product for Cu, CuPd(100) interface and Pd catalysts at different applied potentials. (b) The FE ratios of C2 products to C1 products ($FE_{C2}/FE_{C1}$) at different applied potentials. (c) Current density curves and (d) Tafel slopes for Cu, CuPd(100) interface, and Pd catalysts.

To assess the catalytic activity of the CuPd(100) interface catalyst, the $CO_2$ electroreduction test was carried out. As shown in Fig. 5a, for Cu catalyst, the FE of C2 products gradually increases from $3.7\pm0.4\%$ to $23.6\pm1.5\%$ at the cathode potential range of $-0.8$ $V_{RHE}$ to $-1.4$ $V_{RHE}$, and the FE of C1 products accordingly decreases from $54.7\pm1.3\%$ to $30.6\pm1.4\%$ in the same potential range. For the CuPd(100) interface catalyst, the FE of C2 products increases from $7\pm0.6\%$ to $50.3\pm1.2\%$ at the corresponding potentials, which is 2.1 times higher than that for Cu catalyst ($23.6\pm1.5\%$). For the Pd catalyst, only C1 products and $H_2$ can be detected, and the FE of $H_2$ gradually increases as the cathode potential becomes more negative. More detailed data about the FE of the products for these three catalysts are shown in Fig. S8.

In order to analyse the selectivity of C2 products, Fig. 5b shows the FE ratios of C2 products to C1 products ($FE_{C2}/FE_{C1}$) at different applied potentials. At all the applied potentials, the $FE_{C2}/FE_{C1}$ of CuPd(100) interface catalyst is larger than that of Cu catalyst. Especially at $-1.4$ $V_{RHE}$, the $FE_{C2}/FE_{C1}$ of CuPd(100) interface catalyst reaches 2.4, while for Cu catalyst it is 0.77. This result proved higher selectivity of C2 products on CuPd(100) interface catalyst than that of the Cu catalyst. As previously mentioned, no C2 products can be detected in the Pd catalyst.

Fig. 5c shows the current density at different cathode potentials. The current density of CuPd(100) interface catalyst is significantly greater than that of Cu and Pd catalysts at each potential, indicating a faster reaction kinetics of the CuPd(100) interface catalyst. This performance is also revealed by linear sweep voltammetry (LSV) curves (Fig. S9). Moreover, the CuPd(100) interface catalyst displays the lowest Tafel slope (374 mV dec$^{-1}$) compared to the Cu catalyst (384 mV dec$^{-1}$) and Pd catalyst (467 mV dec$^{-1}$) (Fig. 5d), demonstrating the fast C2 products generation kinetics on CuPd(100) interface.

Furthermore, the electrochemically active surface area (ECSA) tests show the CuPd(100) interface catalyst possess the highest ECSA ($7.04*10^{-3}$ mF cm$^{-2}$), followed by the Cu catalyst ($1.84*10^{-3}$ mF cm$^{-2}$), and the Pd catalyst is the smallest ($0.99*10^{-3}$ mF cm$^{-2}$) (Fig. S10). The electrochemical impedance spectra (EIS) shows that the CuPd(100) interface catalyst has the best electrical conductivity among these three catalysts (Fig. S11). These above results prove that the CuPd(100) interface catalyst has greater activity and selectivity toward C2 products than both Cu catalyst and Pd catalyst.

## 4. Conclusions

In summary, we predicted by DFT calculation that a CuPd(100) interface catalyst should have higher efficiency of C2 products in the $CO_2$ electroreduction reaction, compared to Cu or Pd monometallic catalysts. Our calculations show that the CuPd(100) interface catalyst is able to obtain sufficient CO* for C-C coupling by enhancing the $CO_2$ adsorption and decreasing the energy barrier of $CO_2$* hydrogenation step. The PDS energy barrier of $CO_2$ to C2 products on CuPd(100) interface catalyst (0.61 eV) is smaller than that of Cu(100) (0.72 eV). Guided by the theoretical predictions, the CuPd(100) interface catalyst was synthesized by thermal-reduction treatment and followed with an in-situ growth process. The HRTEM clearly shows the CuPd(100) interface. Combined the $CO_2$-TPD results and gas sensor measurements, we verified the enhanced adsorption of $CO_2$ and the decrease in energy barrier of $CO_2$* hydrogenation on CuPd(100) interface. As a result, the CuPd(100) interface catalyst exhibits a C2 FE of 50.3±1.2% at −1.4 $V_{RHE}$ in 0.1 M $KHCO_3$, which is 2.1 times higher than that of Cu catalyst (23.6±1.5%). The consistency between the theoretical and experimental results provides new sights to design better Cu-based electrocatalyst for conversion of $CO_2$ to desired multi-carbon products.

**Conflicts of interest**

There are no conflicts to declare.

**Acknowledgements**

The authors gratefully thank the Natural Science Foundation of China (Grant No. 21872174, 22002189, 51673217 and U1932148), the International Science and Technology Cooperation Program (Grant No. 2017YFE0127800 and 2018YFE0203402), the Hunan Provincial Science and Technology Program (2017XK2026), the Hunan Provincial Natural Science Foundation (2020JJ2041and 2020JJ5691), the Hunan Provincial Science and Technology Plan Project (Grant No. 2017TP1001), the Shenzhen Science and Technology Innovation Project (Grant No. JCYJ20180307151313532), Ministry of Science and Technology, Taiwan (Grant No. MOST108-2113-M-213-006). Emiliano Cortés acknowledges funding and support from the Deutsche Forschungsgemeinschaft (DFG, German Research Foundation) under Germany´s Excellence Strategy (EXC 2089/1 - 390776260), the Bavarian Solar Energies Go Hybrid (SolTech) program and the Center for NanoScience (CeNS), the European Commission for the ERC Starting Grant CATALIGHT (802989).

**Author contributions**

Min Liu, Junwei Fu, Masahiro Miyauchi and Xiaoliang Liu supervised the project. Min Liu and Akira Yamaguchi designed the experiments and analysed the results. Li Zhu synthesized the samples, performed the electrochemical experiments, and analysed the results. Li Zhu and Kang Liu carried out the DFT calculation and wrote the corresponding section. Yiyang Lin, Ying-Rui Lu and Ting-Shan Chan conducted the EXAFS measurements, and Emiliano Cortés analysed the results. Junhua Hu and Hongmei Li carried out the electron microscope measurements. All authors read and commented on the manuscript.

## Notes and references